\documentclass[12pt]{article}

\usepackage[dvips]{graphicx}
\usepackage{epstopdf}
\usepackage{amsmath}

 \usepackage{amssymb}
 \usepackage{amsfonts}
 \usepackage{latexsym}

 \usepackage{graphicx}
 \usepackage{amssymb}
 \usepackage{amsfonts}
 \usepackage{latexsym}

\newcommand{\nn}{\nonumber\\}
\newcommand{\n}[1]{\label{#1}}

\newcommand{\eq}[1]{(\ref{#1})}

\renewcommand{\baselinestretch}{1.2}
\def\beq{\begin{eqnarray}}
\def\eeq{\end{eqnarray}}
\def\ln{\,\mbox{ln}\,}

\def\Tr{\,\mbox{Tr}\,}

\def\al{\alpha}
\def\be{\beta}

\def\de{\delta}
\def\vp{\varepsilon}
\def\ep{\epsilon}

\def\la{\lambda}
\def\na{\nabla}
\def\pa{\partial}

\def\si{\sigma}
\def\om{\omega}

\def\Ga{\Gamma}

\begin{document}

\begin{center}

{\large{\bf One-loop divergences in the Galileon model}}

\vskip 6mm


\textbf{Tib\'{e}rio de Paula Netto}
\footnote{
E-mail: tiberiop@fisica.ufjf.br}
\quad
\textbf{and}
\quad
\textbf{Ilya L. Shapiro}
\footnote{
E-mail: shapiro@fisica.ufjf.br. On leave from Tomsk State
Pedagogical University, Tomsk, Russia.}
\vskip 6mm
Departamento de F\'{\i}sica, \ ICE, \
Universidade Federal de Juiz de Fora
\\ Juiz de Fora, \ 36036-330, \ MG, \ Brazil
\end{center}

\vskip 12mm

\begin{quotation}
\noindent {\large {\it Abstract}}.
\quad
The investigation of UV divergences is a relevant step in better
understanding of a new theory. In this work the one-loop divergences
in the free field sector are obtained for the popular Galileons
model. The calculations are performed by the generalized
Schwinger-DeWitt technique and also by means of Feynman diagrams.
The first method can be directly generalized to curved space, but
here we deal only with the flat-space limit. We show that the UV
completion of the theory includes the $\pi\Box^4\pi$ term. According
to our previous analysis in the case of quantum gravity, this means
that the theory can be modified to become superrenormalizable, but
then its physical spectrum includes two massive ghosts and one
massive scalar with positive kinetic energy. The effective approach
in this theory can be perfectly successful, exactly as in the higher
derivative quantum gravity, and in this case the non-renormalization
theorem for Galileons remains valid in the low-energy region.
\vskip 3mm

\vskip 3mm


\end{quotation}
\vskip 4mm

\section{Introduction}
\label{int}

The Galileons are qualitatively new models of scalar field with
numerous applications. Originally Galileons were introduced in the
context of Dvali-Gabadadze-Porrati model of gravity \cite{DGP} and
attracted a great deal of interest (see, e.g., \cite{G1,G2,G3} and
further references therein). One of the standard motivations to
consider Galileons is that they are assumed to possess some unusual
renormalization properties \cite{LutRorRat-03,NikRat-04}, especially
when treated in the effective quantum field theory framework
\cite{HinTroWes-10,TroHin-11,BurRhamSeeToll-10}.

In short, Galileons are second derivative theories with much more
derivatives in the action. This means that the kinetic term of such
a model has only two derivatives, while the interacting terms have
much more derivatives. As a result, the tree-level propagator of
the theory is free from higher derivative ghosts and, at the same
time, the possible divergences have so many derivatives that the
terms which are present in the classical action, never get
renormalized \cite{LutRorRat-03,NikRat-04}. In this respect, the
Galileon model is interesting to compare with the fourth derivative
quantum gravity (HDQG) theory \cite{Stelle-77} (see also \cite{book}
for detailed introduction) which was further generalized in
\cite{highderi}. The common point between the two theories is the
presence of higher derivatives. In case of HDQG this makes the theory
renormalizable \cite{Stelle-77} or even superrenormalizable
\cite{highderi}. However, there is a price to pay: the particle
spectrum of the theory includes higher derivative ghost
\cite{Stelle-77} or ghosts \cite{highderi}. At the same time,
Galileon model is {\it very much} nonrenormalizable, in a sense
that there are many possible divergences, however all of the possible logarithmically divergent terms have much more derivatives that the
initial classical action, as a result the low-energy sector of the
theory is free from strong UV quantum corrections.

The described scheme of constructing a theory which is
not affected, in the UV limit, by quantum corrections, looks
very attractive, but there is one important point to verify.
If there are higher derivative divergences in the propagator
sector, then the UV completion of the theory actually has
massive ghost. Eliminating this ghost from the spectrum leads
to the possible unitarity breaking, exactly as in the HDQG.
And, exactly as in the HDQG, we can try to formulate the theory
in such a way that the ghost is not generated at relatively low
energies. So, the program of exploring Galileons at quantum level
should be supplemented by direct calculation of the quantum
contributions to the scalar field propagator. Such a calculation,
at the one-loop level, is the subject of the present paper.

In general, the status of the quantum theory essentially depends
on the structure of UV divergences in the free field sector. It may
happen that the quantum theory is non-renormalizable, or it can be
completed to become renormalizable or even superrenormalizable,
like the HDQG model of \cite{highderi}. In the last case the
one-loop contribution may be the only one which is relevant. In
what follows we shall figure out what is the situation, namely
whether the Galileons theory is non-renormalizable, renormalizable
or superrenormalizable and whether it can be unitary in a
strong sense or only as an effective field theory.

The paper is organized as follows.
In Sect. \ref{Gali} we present a very brief description
of the classical action of the model.
In Sect. \ref{divssect} the derivation of one-loop divergences is
performed by means on the generalized Schwinger-DeWitt technique
\cite{bavi85} with expansions originally developed in
\cite{QGwMatter}. The main advantage of this technique is that
it can be also applied to the similar derivation in curved space,
however in the present Letter we restrict our attention only by
the flat-space limit, which is indeed sufficient to answer the
questions formulated above.
In Sect. \ref{diagras}, which was written for better control and
for illustrative purposes, we show how the same result can be
obtained by means of a more conventional Feynman diagrams technique.
In Sect. \ref{con} we present some additional discussions
of the result and draw our conclusions.

\section{Brief description of the model}
\label{Gali}

In a four dimensional space-time there are only five Lagrangians
with single scalar field $\pi$, which are invariant (up to a total
derivative) under the following transformation
\beq
\n{1}
\pi \rightarrow \pi + b_\mu x^\mu + c\,,
\label{Gali-sim}
\eeq
where $c$ and $b_\mu$ are constants. The transformation \eq{1}
is called {\it Galilean} transformation and the field $\pi$
is called {\it Galileon}. These five Lagrangians can be
represented by the following structures:
\beq
{\cal L}_1
&=&
\pi \,,
\nonumber
\\
{\cal L}_2
&=&
\frac{1}{2} \pa_\mu \pi \pa^\mu \pi\,,
\nonumber
\\
{\cal L}_3
&=&  \pa_\mu \pi \pa^\mu \pi \Box \pi \,,
\nonumber
\\
{\cal L}_4
&=&
\frac{1}{2}\pa_\mu \pi \pa^\mu \pi (\Box \pi)^2
- \pa_\mu \pi \pa^\mu \pa^\nu \pi \pa_\nu \pi\Box\pi
- \frac{1}{2} \pa_\mu \pa_\nu \pi \pa^\mu\pa^\nu \pi
\pa_\varrho \pi \pa^\rho  \pi
\nonumber
\\
&+& \pa_\mu \pi \pa^\mu \pa^\nu \pi \pa_\nu
\pa_\varrho \pi \pa^\rho \pi\,,
\nonumber
\\
{\cal L}_5
&=&
\frac{1}{6}\pa_\mu \pi \pa^\mu \pi (\Box \pi)^3
- \frac{1}{2} \pa_\mu \pi  \pa^\mu \pa^\nu
\pi \pa_\nu \pi   \Box \pi
- \frac{1}{2} \pa_\mu \pa_\nu \pi \pa^\mu \pa^\nu \pi \pa_\varrho
\pi \pa^\rho  \pi \Box \pi
\nonumber
\\
&+&
\pa_\mu \pi \pa^\mu \pa^\nu
\pi \pa_\nu \pa_\varrho \pi \pa^\rho \pi \Box \pi
+ \frac{1}{3}\, \pa_\mu  \pa^\nu \pi \pa_\nu \pa^\varrho
\pi \pa_\varrho  \pa^\mu  \pi \pa_\la  \pi \pa^\la \pi
\nonumber
\\
&+&
\frac{1}{2} \pa_\mu \pa_\nu  \pi \pa^\mu \pa^\nu
\pi \pa_\varrho  \pi  \pa^\varrho  \pa^\la
\pi \pa_\la  \pi - \pa_\mu \pi \pa^\mu \pa^\nu
\pi \pa_\nu \pa_\varrho  \pi \pa^\varrho  \pa^\la
\pa_\la \pi \,.
\label{Lags}
\eeq
The full Lagrangian for the field $\pi$ is a linear combination
of the above Lagrangians
\beq \n{2}
{\cal L}_\pi\,=\, \sum _{i=1}^5 c_i{\cal L}_i\,,
\eeq
where $c_i$'s are generic coefficients.

\section{Calculation of the one-loop counterterms}
\label{divssect}

In this section we shall present the details of the calculation
of the one-loop counterterms of the theory \eq{2}. For the purpose
of calculating the divergences we shall apply the background
field method (see, e.g., Chapter 2 of \cite{book} for introduction)
and the generalized Schwinger-DeWitt technique \cite{bavi85}. Let
us start with the usual splitting of the fields into background
and quantum part
\beq
\n{3}
\pi\rightarrow \pi'=\pi+\sigma.
\eeq
The one-loop effective action is given by the expression
\beq \n{4}
\Gamma^{(1)} &=& \frac{i}{2}\, \Tr \ln \hat{H}\,,
\eeq
where $\hat{H}$ is the bilinear form of the action given by
Lagrangian \eq{2}. Substituting \eq{3} in \eq{2} one can find
the bilinear form of the action
\beq
S^{(2)}=-\frac{1}{2} \int d^4x \,\,\si \,\hat {H}\, \si\,,
\eeq
where $\hat {H}$ has a form
\beq
\hat{H} = \Box + \hat {P}_1 + \hat {P}_2 + \cdots,
\eeq
here $\hat {P}_1 \sim {\cal O}(\pi)$,
$\hat {P}_2 \sim {\cal O}(\pi^2)$ and so on. Let us note that
the scheme of calculation which we are going to perform is
designed to get the correction to the propagator, therefore
we do not need to take into account the terms beyond $\,\hat{P}_2$
and the square of the $\hat {P}_1$-term in this and further expansions.

In order to calculate the divergent part of the one-loop
effective action \eq{4}, in the second order in $\pi$, one can
perform the expansion, which is similar to the one which was
previously used in \cite{QGwMatter} (described also in details
in Chapter 8 of the book \cite{book}),
\beq
\n{6}
\Tr \ln  \hat{H}
&=&
\Tr \ln \big ( \Box + \hat {P}_1 + \hat {P}_2 + \cdots \big )
\nonumber
\\
&=& \Tr \ln \Box
+  \Tr \ln \Big( 1 + \hat {P}_1 \frac{1}{\Box}
+ \hat {P}_2 \frac{1}{\Box} + \cdots \Big)
\nonumber
\\
&=&
\Tr \ln \Box
+  \Tr \Big(\hat{P}_1 \frac{1}{\Box}
+ \hat {P}_2 \frac{1}{\Box}
- \frac{1}{2} \hat {P}_1 \frac{1}{\Box}
\hat {P}_1 \frac{1}{\Box} \Big)\, + \,\cdots
\eeq
The omitted terms are ${\cal O}(\pi^3)$ and hence (as we have
already mentioned above) they are actually irrelevant for our
purposes. In order to reduce the amount of calculations we
shall consider only the particular case of the flat background
space-time. Then the only type of universal trace that
does not vanish has the form
\beq
\n{5}
\Tr \,\partial
_{\mu_1} \cdots \pa_{\mu_{2n-4}}\frac{1}{\Box^n}\Big|_{div}
&=& -\,
\frac{2i}{\ep   }\int d^4x\,\,
\frac{g^{(n-2)}_{\mu_1 \cdots \mu_{2n-4}}}{2^{n-2}\,(n-1)!}\,,
\eeq
with $n \ge 2$ and
\beq
g^{(n-2)}_{\mu_1 \cdots \mu_{2n-4}} &=&
g_{\mu_1\mu_2}\,g_{\mu_3\mu_4}\, \cdots g_{\mu_{2n-3}\mu_{2n-4}}
\,+\,\mbox{all permutations}\,.
\nn
\eeq

By using the formula \eq{5} in \eq{6} we can see that the trace
of the $\hat {P}_2 (1/\Box)$-term corresponds to $n=1$ in Eq.
(\ref{5}) and hence it is finite in dimensional regularization.
Therefore, only the last term of equation \eq{6} gives
contribution to the divergences. Then,
\beq
\n{7}
\Tr \ln  \hat{H} \Big|_{div}
&=& -\, \frac{1}{2} \,\Tr\,
\hat{P}_1 \frac{1}{\Box} \hat {P}_1 \frac{1}{\Box}\Big|_{div}\,.
\eeq
To calculate \eq{7} we need an explicit form of $\hat {P}_1$, namely
\beq
\hat {P}_1
&=&
\hat{U}^{\mu\nu}\,\pa_\mu \pa_\nu\,,
\quad \mbox{where} \quad
\hat{U}^{\mu\nu} \,=\,
4c_3\,\Big[(\Box \pi)g^{\mu\nu}-(\pa^\mu \pa^\nu \pi )\Big].
\n{10}
\eeq

The commutations which enable one to reduce the problem to
the universal trace \eq{5} are performed as follows:
\beq
\n{8}
\Tr \, \hat {P}_1 \,\frac{1}{\Box} \,\hat {P}_1
\,\frac{1}{\Box}\Big|_{div}
&=&
\Tr \, \hat{U}^{\al \be}\pa_\al \pa_\be \,
\frac{1}{\Box}
\hat{U}^{\mu\nu}\pa_\mu \pa_\nu \frac{1}{\Box}\Big|_{div}
\nonumber
\\
&=&
\Tr \,\hat{U}^{\al\be}\pa_\al \pa_\be
\left\{ \hat{U}^{\mu\nu}\pa_\mu \pa_\nu \,\frac{1}{\Box^2}
+ \Big[\frac{1}{\Box},
\hat{U}^{\mu\nu}\Big]\pa_\mu \pa_\nu \frac{1}{\Box}\right\}\Big|_{div}.
\eeq
Furthermore, the last commutator can be transformed as
\beq
\n{9}
&&
\Big[\frac{1}{\Box},\hat{U}^{\mu\nu}\Big]\pa_\mu\pa_\nu
\,\frac{1}{\Box}
\,=\, -(\Box \hat{U}^{\mu\nu})\pa_\mu \pa_\nu \frac{1}{\Box^3}
+(\Box^2 \hat{U}^{\mu\nu})\pa_\mu\pa_\nu \frac{1}{\Box^4}
\nonumber
\\
&&+4(\pa^\la \Box \hat{U}^{\mu\nu})
\pa_\la \pa_\mu \pa_\nu \frac{1}{\Box^4}
-12 ( \pa^\la \pa^\tau \Box \hat{U}^{\mu\nu})
\pa_\la \pa_\tau \pa_\mu \pa_\nu \frac{1}{\Box^5}
- 2(\pa^\la \hat{U}^{\mu\nu})\pa_\la\pa_\mu\pa_\nu\frac{1}{\Box^3}
\nonumber
\\
&&
+ 4(\pa^\la \pa^\tau \hat{U}^{\mu\nu})\pa_\la \pa_\tau
\pa_\mu \pa_\nu \frac{1}{\Box^4}
 - 8 (\pa^\rho \pa^\la\pa^\tau \hat{U}^{\mu\nu})
 \pa_\rho \pa_\la \pa_\tau \pa_\nu \pa_\mu \,\frac{1}{\Box^5}
\nonumber
\\
&&
+ \, 16\, (\pa^\om \pa^\rho \pa^\la \pa^\tau \hat{U}^{\mu\nu})\,
\pa_\om \pa_\rho \pa_\la \pa_\tau \pa_\nu\pa_\mu \frac{1}{\Box^6}
+ {\cal O} \left(\frac{1}{l^5}\right).
\nonumber
\eeq
The terms with background dimension (we assume the reader is
familiar with the terminology of \cite{bavi85}) of more than
$1/l^4$ can be safely omitted here because they do not contribute
to divergences. Substituting the equation \eq{9} into \eq{8} we
finally get, after some tedious calculations, the following
expression:
\beq
\n{11}
&&
\Tr \,
\hat{P}_1 \frac{1}{\Box} \hat{P}_1 \,\frac{1}{\Box}
\Big|_{div}
\,=\,
\Tr \Big\{-\hat{U}^{\al \be }(\pa_\al \pa_\be \Box
\hat{U}^{\mu\nu})\pa_\mu \pa_\nu \frac{1}{\Box^3}
+\hat{U}^{\al \be }(\Box^2 \hat{U}^{\mu\nu})
\pa_\al \pa_\be \pa_\mu\pa_\nu \frac{1}{\Box^4}
\nonumber
\\
&&
+ 8\hat{U}^{\al \be }(\pa_\al \pa^\la \Box
\hat{U}^{\mu\nu})\pa_\be \pa_\la \pa_\mu \pa_\nu
\frac{1}{\Box^4} - 12 \hat{U}^{\al \be}
( \pa^\la \pa^\tau \Box \hat{U}^{\mu\nu})
\pa_\al \pa_\be \pa_\la \pa_\tau \pa_\mu \pa_\nu
\frac{1}{\Box^5}
\nonumber
\\
&&
+4\hat{U}^{\al \be }(\pa_\al \pa_\be \pa^\la
\pa^\tau \hat{U}^{\mu\nu})\pa_\la \pa_\tau \pa_\mu
\pa_\nu \frac{1}{\Box^4} - 16 \hat{U}^{\al \be }
(\pa_\al \pa^\rho \pa^\la \pa^\tau \hat{U}^{\mu\nu})
\pa_\be  \pa_\rho \pa_\la \pa_\tau \pa_\nu \pa_\mu
\frac{1}{\Box^5}
\nonumber\\
&&+16 \hat{U}^{\al \be } (\pa^\omega \pa^\rho
\pa^\la \pa^\tau \hat{U}^{\mu\nu})\pa_\al \pa_\be
\pa_\omega \pa_\rho \pa_\la \pa_\tau \pa_\nu\pa_\mu
\frac{1}{\Box^6} \Big \}
\Big |_{div} .
\eeq

Now we are in position to use the universal trace \eq{5}.
In this way, after certain work, we obtain the following traces
[here $\ep=(4\pi)^2 (n-4)$ in dimensional regularization]:
\beq
- \,\Tr\,\hat{U}^{\al\be}(\pa_\al \pa_\be
\Box \hat{U}^{\mu\nu})\pa_\mu\pa_\nu \,\frac{1}{\Box^3}\Big|_{div}
\,=\, \frac{i}{2\ep} \,
\int d^4x \,\,\hat{U}^{\al\be}
\big(\pa_\al \pa_\be \Box \hat{U}^\mu_\mu\big)\,,
\nonumber
\eeq
\beq
\Tr\,\hat{U}^{\al\be}(\Box^2 \hat{U}^{\mu\nu})\pa_\al\pa_\be
\pa_\mu\pa_\nu\, \frac{1}{\Box^4}\Big|_{div}
\,=\,-\,\frac{i}{12\,\ep}  \int d^ 4x \,\left\{
2\,\hat{U}_{\mu \nu } (\Box^2 \hat{U}^{\mu\nu})
+ \hat{U}^\al_\al (\Box^2 \hat{U}^\mu_\mu) \right\}\,,
\nonumber
\eeq
\beq
&&
8\Tr\,\hat{U}^{\al\be}(\pa_\al \pa^\la \Box \hat{U}^{\mu\nu})
\pa_\be \pa_\la \pa_\mu \pa_\nu \frac{1}{\Box^4}\Big|_{div}
\nonumber
\\
&& =\,
-\frac{2i}{3\ep} \int d^4x \,
\left\{2\hat{U}^\al_\nu (\pa_\al \pa_\mu \Box \hat{U}^{\mu\nu})
+ \hat{U}^{\al\be}(\pa_\al \pa_\be \Box \hat{U}^\mu_\mu)\right\}\,,
\nonumber
\eeq
\beq
&&
4 \Tr\, \hat{U}^{\al \be }(\pa_\al \pa_\be \pa^\la
\pa^\tau \hat{U}^{\mu\nu})\pa_\la \pa_\tau \pa_\mu \pa_\nu
\,\frac{1}{\Box^4}\Big|_{div}
\nonumber
\\
&&
=
\,-\,\frac{i}{3\ep}\int d^4x \,
\Big\{ 2\,\hat{U}^{\al \be}
(\pa_\al \pa_\be \pa_\mu \pa_\nu \hat{U}^{\mu\nu})
+ \hat{U}^{\al \be}(\pa_\al \pa_\be \Box \hat{U}^\mu _\mu) \Big\}\,,
\nonumber
\eeq
\beq
&&
- \, 12 \Tr \hat{U}^{\al \be} (\pa^\la \pa^\tau \Box \hat{U}^{\mu\nu})
\pa_\al \pa_\be \pa_\la \pa_\tau \pa_\mu \pa_\nu
\,\frac{1}{\Box^5}\Big|_{div}
\nonumber
\\
&& =
\frac{i}{4\epsilon}
\int d^4x \,
\Big\{ \hat{U}^{\al \be} (\pa_\al \pa_\be \Box \hat{U}^\mu _\mu )
+ \frac{1}{2}\,\hat{U}^\al _\al (\Box^2 \hat{U} ^\mu_\mu)
+ 4\hat{U}^\al _\nu (\pa_\al \pa_\mu \Box \hat{U}^{\mu\nu})
\nonumber
\\
&& +\,
\hat{U}_{\mu\nu}(\Box^2 \hat{U}^{\mu\nu})
+\hat{U}^\al _\al (\pa_\mu \pa_\nu \Box \hat{U}^{\mu\nu})\Big\}\,,
\nonumber
\eeq
\beq
&&
-\, 16 \Tr \hat{U}^{\al \be }
(\pa_\al \pa^\rho \pa^\la \pa^\tau \hat{U}^{\mu\nu})
\pa_\be \pa_\rho \pa_\la \pa_\tau \pa_\nu \pa_\mu
\,\frac{1}{\Box^5}\Big|_{div}
\nonumber
\\
&& =\,
\frac{i}{2\epsilon}
\int d^4x \,
\left\{ \hat{U}^{\al\be} (\pa_\al \pa_\be \Box \hat{U}^{\mu}_\mu)
+ 2 \hat{U}^{\al \be } (\pa_\al \pa_\be \pa_\mu \pa_\nu \hat{U}^{\mu\nu})
+ 2 \hat{U}^\al_\nu(\pa_\al \pa_\mu \Box \hat{U}^{\mu\nu}) \right\}\,,
\nonumber
\eeq
and
\beq
&&
16 \,\Tr\, \hat{U}^{\al\be}
(\pa^\om \pa^\rho \pa^\la \pa^\tau \hat{U}^{\mu\nu})\,
\pa_\al\pa_\be\pa_\om \pa_\rho \pa_\la\pa_\tau \pa_\nu\pa_\mu
\,\frac{1}{\Box^6}\Big|_{div}
\nonumber
\\
&& =\,-\,
\frac{i}{5\,\ep}
\int d^ 4x \,
\Big\{ \frac{1}{4}\,\hat{U}^\al_\al
\big(\Box^2 \hat{U}^\mu _\mu\big)
+ \hat{U}^\al_\al \big(\pa_\mu \pa_\nu \Box \hat{U}^{\mu\nu}\big)
+ \hat{U}^{\al \be}\big(\pa_\al  \pa_\be \Box \hat{U}^\mu_\mu\big)
\nonumber
\\
&& +\,
2\hat{U}^{\al\be}(\pa_\al\pa_\be\pa_\mu\pa_\nu \hat{U}^{\mu\nu})
+ 4\hat{U}^\al_\nu \big(\pa_\al \pa_\mu \Box \hat{U}^{\mu\nu}\big)
+\frac{1}{2}\,
\hat{U}_{\mu\nu}\big(\Box^2 \hat{U}^{\mu\nu}\big)\Big\}.
\label{traces}
\eeq

By using the traces listed above, Eq. \eq{11} can be reduced to
\beq
&&
\Tr\, \hat {P}_1 \,\frac{1}{\Box} \,\hat {P}_1 \,
\frac{1}{\Box}\Big|_{div}\,
=\,\frac{2i}{\epsilon  } \int d^4x \,
\Big\{\,\frac{1}{40}
\,\hat{U}^{\al\be}(\pa_\al \pa_\be\Box \hat{U}^\mu_\mu)
-\frac{1}{120}\,\hat{U}_{\mu\nu} (\Box^2 \hat{U}^{\mu\nu})
\nonumber
\\
&& -\
\frac{1}{240}\,\hat{U}_\al ^\al  (\Box^2 \hat{U}^\mu_ \mu)
-\frac{1}{30}\hat{U}^{\al\be}(\pa_\al \pa_\be \pa_\mu \pa_\nu
\hat{U}^{\mu\nu})
-\frac{1}{15}\,\hat{U}^\al _\nu (\pa_\al \pa_\mu \Box \hat{U}^{\mu\nu})
\nonumber
\\
&& +\,
\frac{1}{40}\,\hat{U}^\al_\al
(\pa_\mu\pa_\nu\Box\hat{U}^{\mu\nu})\,\Big\}.
\label{P1P1}
\eeq
Replacing, term by term, the explicit form of operator
$\hat{U}^{\mu\nu}$ given by Eq. \eq{10} into Eq. (\ref{P1P1}),
after some algebra we arrive at
\beq
\Tr\,  \hat {P}_1 \,\frac{1}{\Box}\,
 \hat {P}_1 \,\frac{1}{\Box}\Big|_{div}
\,=\,-\,\frac{2i}{\vp}\,c_3^2\,\int d^4x \,\, \pi \Box^4 \pi\,.
\eeq
Finally, from this expression and the equations \eq{4}, \eq{7} we
obtain the result for the one-loop divergences of effective action
in the free field sector,
\beq
\Ga^{(1)}_{div}
&=& -\,
\frac{c_3^2}{2\,\ep} \,\int d^4 x \,\,\pi \Box^4 \pi\,.
\label{divs}
\eeq

The expression (\ref{divs}) is providing us some
relevant information about the quantum properties of the
Galileons theory. Let us present it in the systematic form.
\vskip 2mm

\noindent
{\large $\bullet$} \ The UV logarithmic divergences of the
theory require \ $\pi \Box^4 \pi$-type counterterms. This
means that the consistency of the theory requires that the
same term is included into the classical action of the theory.
If we do not include such a term, it will emerge with infinite
coefficient anyway and then no control over this term via the
effective approach will be possible.
\vskip 2mm

\noindent
{\large $\bullet$} \ According to the analysis of the similar
gravitational theory in \cite{highderi}, the theory with an
extra \ $\pi \Box^4 \pi$-type classical term has, typically,
two massive ghosts (excitations with negative kinetic energy),
one massive scalar degree of freedom and, of course, the
``original'' massless scalar mode. This situation means that
the propagator of the field $\pi$ has the general structure
\beq
G(p) &=&
\frac{1}{p^2} \,-\,\frac{A_1}{p^2+m_1^2}
\,+\,\frac{A_2}{p^2+m_2^2} \,-\,\frac{A_3}{p^2+m_3^2}\,,
\label{propa}
\\
&& A_{1,2,3} > 0\,, \qquad m_3 > m_2 > m_1 > 0\,.
\nonumber
\eeq
Here we assumed Euclidean signature and that there are
no tachyons in the spectrum. The last can be always
provided by adjusting the coefficients of subleading
$\pi \Box^2 \pi$-type and $\pi \Box^3 \pi$-type terms.
Let us note that similar, curvature-dependent, terms are
likely to be requested also by the divergences in curved
space-time. On the other hand, all considerations presented
below should be valid also in the presence of tachyons.
\vskip 2mm

\noindent
{\large $\bullet$} \ According to the analysis of the
similar gravitational theory in \cite{highderi}, the
theory with an extra \ $\pi \Box^4 \pi$-type classical
term is superrenormalizable, for any choice of the
coefficients $c_k$ in the interacting sector of (\ref{Lags}).
This means, in our case, that the divergences can emerge
only at the one-loop level. Starting from the second loop
only one-loop sub-diagrams can be divergent. This feature
does not depend on the flatness of space-time and will
definitely hold also in curved space-time case (see
\cite{book} for introduction to the general theory of
renormalization in curved space-time).
\vskip 2mm

\noindent
{\large $\bullet$} \
The effective approach in the quantum Galileon theory is
perfectly possible if we put a sufficiently small parameter
in front of the  classical \ $\int \pi \Box^4 \pi$-term
in the action. Such parameter should have the form of
$M^{-6}$ and, therefore, the choice of a small coefficient
of the higher derivative term means we choose a huge mass
parameter $M$. The masses of both massive ghosts $m_1$, $m_3$
and of the massive scalar with positive kinetic energy, $m_2$,
will be of the same order of magnitude as $M$ (see \cite{highderi}
for details). If we consider classical or quantum phenomena
at the energies much smaller than $M$, the massive modes
of the scalar do not become active and the conclusions of
\cite{LutRorRat-03,NikRat-04} and
\cite{HinTroWes-10,TroHin-11} remain correct,
including the non-renormalization theorem. In the effective
framework the ``correct'' quantum result is supposed to
be the one we have derived above, and not the one of
the complete theory with the \ $\int \pi \Box^4 \pi$-term.
Finally, this means that (\ref{Gali-sim}) is the low-energy
symmetry, so the fact that it is violated by the one-loop
divergence (\ref{divs}) is irrelevant in this framework.

\section{One-loop divergences from Feynman diagrams}
\label{diagras}

In the previous Section we calculated the one-loop logarithmic
divergences by means of the generalized Schwinger-DeWitt technique.
This approach has an advantage because it is relatively easy to
generalize the results for a curved space-time. However, since in
practise we are dealing only with flat space-time limit, here we
perform an equivalent calculations by a more traditional Feynman
diagrams-based calculation. The purpose of this section is to
have an extra control of the result and also for the illustrative
reasons. Since the theory looks unusual, this consideration may
be instructive. Without going into full details, we will also
provide a comparison between diagrams and universal traces,
considered in the previous Section.

Let us considerer the diagrams which contribute to the two-point
function of the Galileon field at the one-loop level. The first
set of diagrams which are all generated by the Lagrangian
${\cal L}_3$ is shown in Fig. 1.
\vskip 8mm

\noindent
 \begin{tabular}{c}
 \includegraphics[scale=0.94]{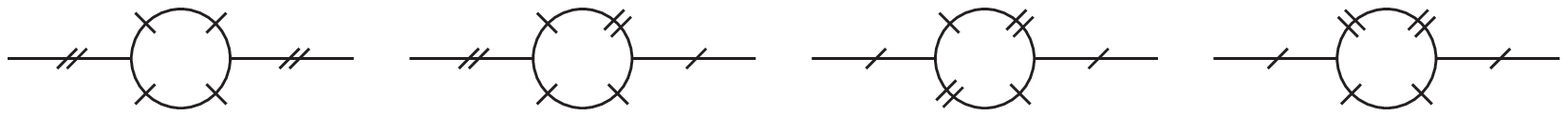}
 \end{tabular}
\begin{quotation}
{\bf Figure 1.} \ The first set of diagrams coming from the
${\cal L}_3$-term vertex, which contribute to the propagator
of the Galileon field. Here primes indicate derivatives.
\end{quotation}

We will be interested in the behavior in the large momentum
regime. The integral associated with the first diagram from
Fig. 1 is
\beq
\Pi _1 (p) = c_3^2
\int \frac{d^4 q}{(2 \pi)^4} \,\,
\frac{p^4 q_\mu q_\nu (q^\mu -p^\mu)( q^\nu -p^\nu)}
{q^2(q-p)^2} \,.
\n{pol1}
\eeq
Expanding the denominator of equation \eq{pol1}, we found at large
momentum scale the relation (in this equation we omit all tensorial
indices and coefficients, and are only interested in power counting
related to the divergences and powers of momentum \ $p$)
\beq
\Pi _1 (p)\,\, \stackrel{q  \rightarrow  \infty }{\sim} \,\,
c_3^2 \int _0 ^\infty dq \,
\Big( \cdots + p^6 q + p^7 + \frac{p^8}{q} + \cdots \Big) \,.
\eeq
One can easily see that this diagram contains ultraviolet
logarithmic divergences and the power of \ $p$ \ corresponding to
this divergence is eight, which fits the \ $\Box^4$-term obtained
in the previous Section.

The others diagrams of our interest are the ones shown in
Fig. 2

\noindent
\vskip 4mm
 \begin{tabular}{c}
 \includegraphics[scale=1]{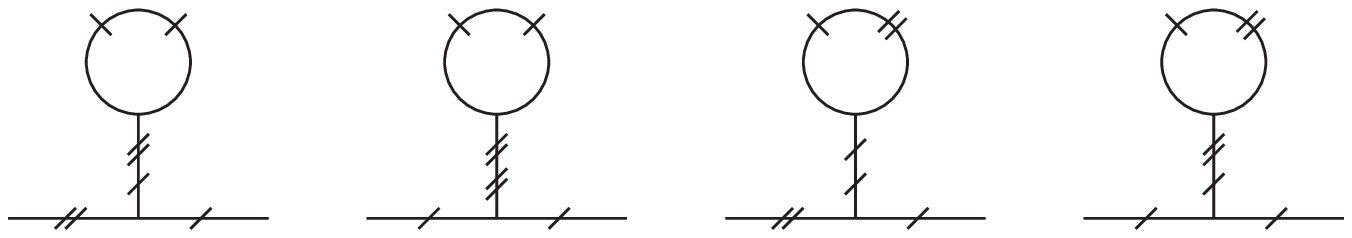}
 \end{tabular}
\begin{quotation}
{\bf Figure 2.} \ The second set of (tadpole-like) diagram provided
by ${\cal L}_3$, which contribute to the two-point function.
\end{quotation}
and also the graphs shown in Fig. 3,
\vskip 8mm
\noindent
 \begin{tabular}{c}
 \mbox{\hspace{+1.9cm}}
 \includegraphics[scale=1]{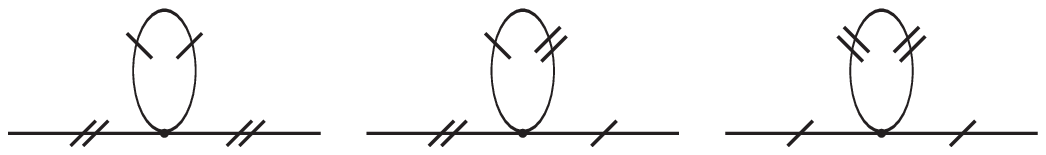}
 \end{tabular}
\begin{quotation}
{\bf Figure 3.} \ The diagrams generated by
lagrangian ${\cal L}_4$ which may contribute for propagator of
Galileon Model.
\end{quotation}
\vskip 1mm

In fact, the diagrams from Figures 2 and 3 do not contribute
to the divergences in the case of Galileon theory. The reason
is that these diagrams include derivatives of the propagator
in a single space-time point, and hence vanish.

Using this simple analysis of the diagrams one can explain
qualitatively the contributions for the divergences for each
term in the expansion for effective action (\ref{6}). First let
us consider the term \ $\Tr \hat{P_1}\frac{1}{\Box}$. \
Since $\hat{P}_1$ is ${\cal O}(\pi^3)$, it contains only
contributions of the Lagrangian ${\cal L}_3$ and is proportional
to $c_3$. Then, $\Tr \hat{P_1}\frac{1}{\Box}$ is also
proportional to $c_3$ and we can see that this term does not
contribute to divergences because it corresponds to Feynman
diagrams which are proportional to $c_3$ and are zero after we
use Wick's theorem, since ${\cal L}_3$ has an odd number of fields.

The next term is \ $\Tr \hat{P_2}\frac{1}{\Box}$. Remember that
$\hat{P}_2$  is ${\cal O}(\pi^4)$ and is proportional to $c_4$,
then the contribution of this term are given by tadpoles diagrams
of Fig. 3. As we have mentioned above, this diagrams do not
contribute, hence we can see why this term makes no contribution
to the counterterms. Finally, the last trace is
\ $\Tr \hat{P_1}\frac{1}{\Box}\hat{P_1}\frac{1}{\Box}$.
This term is proportional to $c_3^2$, its contribution to the
logarithmic divergences is different from zero and is given by
diagrams of Fig. 1.

The considerations presented  above enable us to write the
expression for the divergent part of the two-point function
as presented in Fig. 4.

\vskip 4mm
\noindent
 \begin{tabular}{c}
 \includegraphics[scale=1.2]{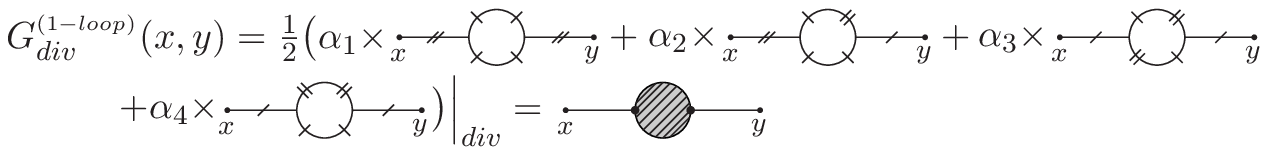}
 \end{tabular}
\begin{quotation}
{\bf Figure 4.} \ Diagrammatic representation of the
Green function. $\al_{1,2,3,4}$ are combinatorial coefficients.
\end{quotation}
\vskip 4mm
\noindent
The definition of the full polarization operator is given
in Fig. 5.
\vskip 4mm
 \begin{tabular}{c}
 \mbox{\hspace{+3.7cm}}
 \includegraphics[scale=1]{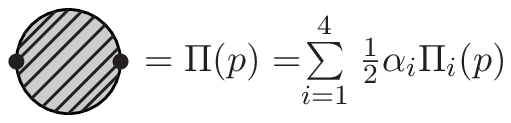}
 \end{tabular}
\begin{quotation}
{\bf Figure 5.} Full polarization operator.
\end{quotation}
\noindent
with
\beq
\Pi_1 (p) = c_3^2 \int \frac{d^4 q}{(2 \pi)^4} \,
\frac{p^4 q_\mu q_\nu (q^\mu -p^\mu)
( q^\nu -p^\nu)}{q^2(q-p)^2} \,,
\eeq
\beq
\Pi _2 (p) = c_3^2
\int \frac{d^4 q}{(2 \pi)^4} \,
\frac{p^2 q^2 q_\mu p_\nu (q^\mu -p^\mu)
( q^\nu -p^\nu)}{q^2(q-p)^2} \,,
\eeq
\beq
\Pi_3 (p) = c_3^2
\int \frac{d^4 q}{(2 \pi)^4} \,
\frac{p_\mu p_\nu q^2 q^\mu ( p^\nu -q^\nu) (p_\al -q_\al )
(p^\al -q^\al)}{q^2(q-p)^2}
\eeq
and
\beq
\Pi_4 (p) = c_3^2
\int \frac{d^4 q}{(2 \pi)^4} \,
\frac{p_\mu p_\nu q^4 ( p^\mu -q^\mu)
( p^\nu -q^\nu) }{q^2(q-p)^2} \,.
\eeq

To evaluate these integrals we used dimensional regularization,
reformulating the theory in the space-time of $2\om$ complex
dimensions, where the integrals are convergent. Using the
formulas of the Appendix, we found the result
\beq
\Pi_1 (p,\om) &=&  \frac{ i\, c_3 ^2}{4} \,p^8 \,I_1 \,,
\nonumber
\\
\Pi_i (p,\om ) &=&  0 \,, \qquad i=2,3,4\,.
\label{Pis}
\eeq
The integral $I_1$ is defined in the Appendix. To find the
divergent part of the polarization operator we consider
the limit $\om \rightarrow 2$. The divergent part is given
by the pole of \ $\Gamma (2-\om)$ \ in \ $I_1$. We find
\beq
I_1 \big|_{div} = -\frac{2}{\epsilon },
\eeq
where $\epsilon = (4\pi)^2(n-4)$. Taking into account the
combinatorial coefficient $\al_1=4$, we arrive at
\beq
\Pi_{div}(p,\om) \,=\, \frac{i\, c_3 ^2}{2}\, p^8\, I_1^{div}
\,=\,  -\,\frac{i\, c_3^2}{\epsilon}\, p^8 \,.
\eeq

The last part is to calculate the diveregent part of Effective
Action in the coordinate representation. For this end one can use
the Dyson's formula
\beq
G^{-1}\,=\,G^{-1}_0\,-\,\Pi(p)\,.
\label{Dyson-gen}
\eeq
Then the divergent part of Effective Action can be written as
\beq
\n{dyson}
\frac{\de ^2 \Ga_{div}^{(1)}}{\de \pi \de \pi}
\,=\, -\, i \Pi_{div}(\Box)
\, =\, - \,\frac{c_3^2}{\epsilon}\,\Box^4 \,.
\eeq
The equation \eq{dyson} can be easily solved and we arrive at
\beq
\Ga^{(1)}_{div}
&=& -\,\frac{c_3^2}{2\,\ep} \,\int d^4 x \,\,\pi \Box^4 \pi\,,
\eeq
that is exactly the result obtained in previous section.

\section{Concluding discussions}
\label{con}

We have developed the background field method and
calculated the one-loop divergences for the Galileon model.
It turns out that the UV completion of the theory includes
higher derivative sectors, as it was indeed anticipated in
\cite{LutRorRat-03,NikRat-04}. An interesting new aspect is
that this UV completion leads to the superrenormalizable
quantum theory, where only the one-loop contribution to the
effective action is divergent and everything beyond the
one-loop order is finite.

There is an interesting similarity between the quantum
Galileons model with this higher derivative completion and
the higher derivative quantum gravity (HDQG). In fact, the
unique conceptual difference is that the Galileon model
with an extra  \ $\int \pi \Box^4 \pi$-term is strongly
superrenormalizable, while HDQG admits different levels
of renormalizable and superrenormalizable theories.
At the same time, the status of ghosts in these two theories
is very close. In case of HDQG the Planck mass $M_P$ plays the
role of the massive parameter $M$ which was discussed at
the end of the section \ref{divssect}. In both cases one
can provide the absence of ghosts at the tree level for
sufficiently low energies.

One simple test of the last statement has been applied
recently in the cosmological framework \cite{GW-Stab}.
It was shown that the physically relevant cosmological
solutions in the higher derivative gravity theory (even with
complicated semiclassical corrections) are stable with respect
to graviton perturbations (gravitational waves). Definitely,
this output is expected to hold only until we do not start to
deal with the perturbations with the initial amplitude of the
Planck order of magnitude. However, after the Universe passed
through its initial Planck-scale epoch, such violent
perturbations are never generated, and therefore the theory
is safe at the classical level. Probably, this should mean
that the quantum theory is also free of the ghost problem at
the tree level. Of course, this is not an obvious statement,
because it is not really clear how the most relevant classical
solutions (such as cosmological one, for example) of the
gravitational theory can be reproduced via the linearized
gravity approach. At the same time, the stability of the
cosmological solution \cite{GW-Stab} is definitely more
fundamental issue than our skills in linearizing gravity,
so we can definitely say that we have a strong positive
arguments in favor of higher derivative
theories\footnote{There are many other interesting
proposals towards the solution of the ghost issue in HDQG
\cite{tom,anttom,Hawking,LeoMod,Riv} which can be also
productive. In any case it is important to care about
higher derivative terms in gravity, since they are
requested by consistency of the quantum theory of
matter fields \cite{book,PoImpo}.}.

Let us, finally, discuss some practical lessons which we can
learn from the analogy with the HDQG case. As far as the main
applications of Galileons is related to cosmology, it would be
definitely interesting to consider the stability of the
classical cosmological solutions in the presence of higher
derivative terms which are a necessary UV completion of the
theory. For this end one has to complete the
 \ $\int \pi \Box^4 \pi$-term derived here by the
corresponding curvature-dependent terms. Regardless of most
of the relevant information for such a cosmological application
can be perfectly well obtained from power counting arguments,
it would be anyway reasonable to generalize our calculation
of the one-loop divergences to the curved space case.

\section{An extra note on more general scalar-tensor theories}
\label{Duzia}

It is possible, in principle, to generalize the results
considered above to the more complicated and general
scalar-tensor model with second-order field equations.
Such a theory has been recently considered in \cite{FKT}
(see further references therein).
The action of this general model does not satisfy the
symmetry (\ref{Gali-sim}) and this opens the way for
infinitely many new terms in the Lagrangian. The form
of the curved-space Lagrangian is
\beq
{\cal L} &=& \sum\limits_{i=2}^{5}
{\cal L}_i\,,
\label{Lags-FKT}
\\
{\cal L}_2
&=& K(\phi,X)\,,\qquad \mbox{where}\quad
X=-\frac{1}{2} \pa_\mu \phi \pa^\mu \phi\,,
\nonumber
\\
{\cal L}_3
&=& -\,G_3(\phi,X)\Box \phi \,,
\nonumber
\\
{\cal L}_4
&=&
G_4(\phi,X)R\, + \,G_{4,X}\big[(\Box \phi)^2
- (\na_\mu\na_\nu\phi)(\na^\mu\na^\nu\phi)\big]
\nonumber
\\
{\cal L}_5
&=&
G_5(\phi,X)G_{\mu\nu}
-\,\frac16\,G_{5,X}
\big[(\Box \phi)^2
- 3(\Box \phi)(\na_\mu\na_\nu\phi)(\na^\mu\na^\nu\phi)
\nonumber
\\
&&
+ 2(\Box \phi)(\na^\mu\na_\al\phi)(\na^\al\na_\be\phi)
(\na^\be\na_\mu\phi)\big]\,,
\nonumber
\eeq
where $G_k(\phi,X)$, with $k=3,4,5$, are arbitrary functions
and $G_{k,X}(\phi,X)$ are the corresponding derivatives
with respect to $X$.

The question in which we are interested in is whether and
how the result (\ref{divs}) gets modified in the more
general model (\ref{Lags-FKT}). In order to address this issue,
it is sufficient to perform the analysis of the power counting.
Hence, our consideration of this model will be brief,
so that we leave the details of the performed analysis as
exercise for an interested reader and give only the main
result.

As far as we are interested only in the counterterms
which contribute to the propagator of $\phi$ and do not
intend to quantize metric, the modifications which come
from the curvature-dependent terms in (\ref{Lags-FKT}) are
irrelevant. Furthermore, according to our previous
analysis, only the contributions to the vertices with
three legs are significant, while the vertices with
four and more legs play no role. Taking these two
observations into account, one can easily see that
the result strongly depends on the presence of a
constant term in the function $G_{5,X}$.

Let us assume that $G_5(\phi,X)$ can be expanded as
\beq
G_5(\phi,X) = G_{50}(\phi) + G_{51}(\phi)X +
G_{52}(\phi)X^2 + \,...\,,
\nonumber
\\
\mbox{where}\qquad
G_{51}(\phi)X = G_{510} + G_{511}\,\phi
+ G_{512}\,\phi^2 + G_{513}\,\phi^3 + \,...\,.
\label{novje}
\eeq
It is easy to see that if the coefficient $G_{510}$ in
(\ref{novje}) is zero, then the result  (\ref{divs})
does not change (except the coefficient, of course).
However, in case \ $G_{510}\neq 0$ \ the consideration
of the superficial degree of divergences of the
relevant diagram shows that the leading counterterm
will be different from (\ref{divs}). In this case we
can expect the counterterms of the form
\beq
\Ga^{(1)}_{div}
&\sim &
\int d^4 x \,\,\phi \big( \Box^6 + ...\big)\phi\,,
\label{divs-12}
\eeq
where the dots indicate the presence of possible terms
with lower powers of $\Box$. The higher order in
derivatives in (\ref{divs-12}) compared to (\ref{divs})
is because the  $G_{510}$-term in (\ref{Lags-FKT})
leads to the vertices with three legs and five
derivatives, which are not present in the Galileon
case (\ref{Lags}). Qualitatively, the consequences of
the higher derivative terms remain the same as it was
discussed above. The theory with the corresponding
UV completion would be superrenormalizable, and possesses
a (larger, in this case) set of ghosts and massive
scalar states with positive kinetic energy.

\section*{Appendix. Massless integrals}

To calculate the integrals of Feynman diagrams from Sect. 4, we
need the divergent parts of some massless integrals, which are
given below. The basic formulas \eq{I1}-\eq{I5} can be found in
\cite{Leib} and other integrals can be obtained by the method
explained in \cite{Capper}. All integrals are defined over
Euclidian space and must be understood though the prescription
for massless case explained in \cite{Leib}
\beq \n{I1}
\int \frac{d^{2 \om}q}
{(2 \pi)^{2 \om} q^2(q-p)^2}=I_1 \,,
\eeq
\beq
\int \frac{d^{2 \om}q \, q_\mu}
{(2 \pi)^{2 \om} q^2(q-p)^2}=p_\mu I_1 \,,
\eeq
\beq
\int \frac{d^{2 \om}q \, q_\mu q_\nu}
{(2 \pi)^{2 \om} q^2(q-p)^2}= \de_{\mu \nu} I_3 +
p_\mu p_\nu I_4 \,,
\eeq
\beq
\int \frac{d^{2 \om}q \, q_\mu q_\nu q_\al}
{(2 \pi)^{2 \om} q^2(q-p)^2}= p_\mu p_\nu p_\alpha I_5
+ E_{\mu \nu \al} I_6\,,
\eeq
\beq \n{I5}
\int \frac{d^{2 \om}q \, q_\mu q_\nu q_\al q_\be}
{(2 \pi)^{2 \om} q^2(q-p)^2}= p_\mu p_\nu p_\alpha p_\be I_7
+ G_{\mu \nu \al \be} I_8 + H_{\mu \nu \al \be} I_9 \,,
\eeq
\beq
\int \frac{d^{2 \om}q \, q_\mu q_\nu q_\al q_\be q_\rho}
{(2 \pi)^{2 \om} q^2(q-p)^2}=
p_\mu p_\nu p_\alpha p_\be p_\rho I_{10}
+ K_{\mu \nu \al \be \rho} I_{11}
+ L_{\mu \nu \al \be \rho} I_{12} \,,
\eeq
\beq
\int \frac{d^{2 \om}q \, q_\mu q_\nu q_\al q_\be q_\rho q_\om}
{(2 \pi)^{2 \om} q^2(q-p)^2} &=&
p_\mu p_\nu p_\alpha p_\be p_\rho p_\om I_{13}
+ R_{\mu \nu \al \be \rho \om} I_{14}
+ S_{\mu \nu \al \be \rho \om} I_{15}
\\
&+& T_{\mu \nu \al \be \rho \om} I_{16} \,,
\nonumber
\eeq
where
\beq
E_{\mu \nu \al }= \de_{\mu \nu} p_\al \,
+\,\mbox{all permutations}\,,
\eeq
\beq
G_{\mu \nu \al \be} &=&
\de_{\mu \nu} p_\al p_\be +
\,+\,\mbox{all permutations}\,,
\eeq
\beq
H_{\mu \nu \al \be}= \de_{\mu \nu} \de_{\al \be}
\,+\,\mbox{all permutations}\,,
\eeq
\beq
K_{\mu \nu \al \be \rho}&=&
 \de_{\mu \nu} p_\al p_\be p_\rho
\,+\,\mbox{all permutations}\,,
\eeq
\beq
L_{\mu \nu \al \be \rho}&=&
\de_{\mu \nu} \de_{\al \be} p_\rho
\,+\,\mbox{all permutations}\,,
\eeq
\beq
R_{\mu \nu \al \be \rho \om}
&=& \de_{\mu\nu}p_\al p_\be p_\rho p_\om
\,+\,\mbox{all permutations}\,,\eeq
\beq
S_{\mu \nu \al \be \rho \om}
&=& \de_{\mu\nu} \de_{\al \be} p_\rho p_\om
\,+\,\mbox{all permutations}\,,\eeq
\beq
T_{\mu \nu \al \be \rho \om}
&=& \de_{\mu\nu} \de_{\al \be} \de_{ \rho \om}
\,+\,\mbox{all permutations}\,.
\eeq
The integrals $I_2,\,...\,,I_{16}$ can be expressed in terms of
the basic integral $I_1$,
\beq
I_1 \equiv \frac{1}{(4\pi)^\om \Ga( 2\om -2)}
\, \Ga (2-\om) \Ga (\om -1) \Ga (\om -1) p^{2(\om -2)}
\eeq
and are given by the expressions
\beq
I_2=\frac{1}{2}I_1 \,, \qquad
\eeq
\beq
I_3=\frac{-p^2}{4(2\om-1)}I_1 \,,
\eeq
\beq
I_4=\frac{\om}{2(2\om-1)}I_1 \,,
\eeq
\beq
I_5=\frac{(\om+1)}{4(2\om-1)}I_1 \,,
\eeq
\beq
I_6=\frac{-p^2}{8(2\om-1)}I_1 \,,
\eeq
\beq
I_7=\frac{(\om+1)(\om+2)}{4(4\om^2-1)}I_1 \,,
\eeq
\beq
I_8=\frac{-(\om +1)p^2}
{8(4\om^2-1)}I_1 \,,
\eeq
\beq
I_9=\frac{p^4}{16(4\om^2-1)}I_1 \,,
\eeq
\beq
I_{10}=\frac{(\om +3)( \om +2)}
{8(4\om^2-1)}I_1 \,,
\eeq
\beq
I_{11}=\frac{-(\om +2)p^2}{16(4\om^2-1)}I_1 \,,
\eeq
\beq
I_{12}=\frac{p^4}{32(4\om^2-1)}I_1 \,,
\eeq
\beq
I_{13}=\frac{(\om +4)(\om +3)(\om + 2)}
{8(2\om + 3)(4\om^2-1)}I_1 \,,
\eeq
\beq
I_{14}=\frac{-(\om +3)(\om + 2)p^2}
{16(2\om + 3)(4\om^2-1)}I_1 \,,
\eeq
\beq
I_{15}=\frac{(\om + 2)p^4}{32(2\om + 3)(4\om^2-1)}I_1 \,,
\eeq
\beq
I_{16}=\frac{-p^6}{64(2\om + 3)(4\om^2-1)}I_1 \,.
\eeq

\section*{Acknowledgments}
T.P.N. thanks FAPEMIG for supporting his Ms. project. \ I.Sh.
is grateful to CNPq, CAPES, FAPEMIG and ICTP for partial
support of his work.

\renewcommand{\baselinestretch}{0.9}

\end{document}